\documentclass[12pt ]{article}

\renewcommand{\theequation}
{\arabic{section}.\arabic{equation}}

\makeatletter
\def\eqnarray{ \stepcounter{equation} \let\@currentlabel=\theequation
 \global\@eqnswtrue
 \global\@eqcnt\z@
 \tabskip\@centering
 \let\\=\@eqncr
 $$\halign to \displaywidth\bgroup\@eqnsel\hskip\@centering
 $\displaystyle\tabskip\z@{##}$&\global\@eqcnt\@ne
 \hfil$\displaystyle{{}##{}}$\hfil
 &\global\@eqcnt\tw@$\displaystyle\tabskip\z@{##}$\hfil
 \tabskip\@centering&\llap{##}\tabskip\z@\cr}
\makeatother

\makeatletter
\def\@arrayacol{\edef\@preamble{\@preamble \hskip .5\arraycolsep}}
\def\array{\let\@acol\@arrayacol \let\@classz\@arrayclassz
\let\@classiv\@arrayclassiv \let\\\@arraycr\def\@halignto{}\@tabarray}
\makeatother

\makeatletter
\newcounter{subeqncnt}
\def\thesubeqncnt{\alph{subeqncnt}}
\def\subequations{\begingroup%
   \stepcounter{equation}\edef\@tempa{\theequation}%
   \let\c@equation\c@subeqncnt\c@subeqncnt\z@
   \edef\theequation{\@tempa\noexpand\thesubeqncnt}}

\makeatother

\newcommand{\be}{\begin{equation}}
\newcommand{\ee}{\end{equation}}

\newcommand{\beqa}{\begin{eqnarray}}
\newcommand{\eeqa}{\end{eqnarray}}
\newcommand{\nn}{\nonumber}

\newcommand{\eqref}[1]{(\ref{#1})}

\def\CN {{\cal N}}

\begin{document}

\setlength{\baselineskip}{7mm}
\begin{titlepage}
\begin{flushright}

{\tt NRCPS-HE-38-2011} \\
{\tt CERN-PH-TH/2011-175}\\

\end{flushright}

\begin{center}
{\Large ~\\{\it Conformal Invariance of Tensor Boson  Tree Amplitudes\\
\vspace{1cm}

}

}

\vspace{1cm}

{\sl Ignatios Antoniadis$^{1}$\footnote{${}$ On leave of absence
from CPHT {\'E}cole Polytechnique, F-91128, Palaiseau Cedex,France.}
 and  George Savvidy$^{1,3}$

\bigskip
\centerline{${}^1$ \sl Department of Physics, CERN Theory Division CH-1211 Geneva 23, Switzerland}
\bigskip
\centerline{${}^3$ \sl Demokritos National Research Center, Ag. Paraskevi,  Athens, Greece}
\bigskip

}
\end{center}
\vspace{60pt}

\centerline{{\bf Abstract}}
The BCFW recursion relation allows to find out the tree-level scattering amplitudes
for gluons and tensor gauge bosons in generalized Yang-Mills theory. We demonstrate that the corresponding
MHV amplitudes for the tensor gauge bosons of spin-$s$ and $n$ gluons are
invariant under conformal group of transformations. This is highly unexpected result for
the higher spin particles, in particular this
is not true  for the scattering amplitudes of gravitons. We discuss and compare
the  tree-level scattering amplitudes for the charged tensor bosons with the
corresponding scattering amplitudes for gravitons, stressing their differences and similarities.

\vspace{12pt}

\noindent

\end{titlepage}



\pagestyle{plain}

\section{\it Introduction}

The Lagrangian of non-Abelian tensor gauge fields describes the interaction of
the gluons with their massless excitations of higher
spin \cite{Savvidy:2005fi,Savvidy:2010vb,Antoniadis:2011re}. The characteristic property of
generalized Yang-Mills theory is that all interaction vertices
between gluons and their high-spin excitations have {\it dimensionless coupling constants} in four-dimensional space-time.
That is, the cubic interaction vertices have only first order derivatives and the quartic vertices
have no derivatives at all.

For better understanding of the model it is important to study the corresponding tree level
scattering amplitudes. One of the first calculations of tree level amplitudes
was made in the article  \cite{Georgiou:2010mf}
where the authors considered the production of tensor gluons in the annihilation
processes of quarks and gluons.
It uses a very powerful spinor helicity technique and  color
decomposition for the calculation of high order tree level diagrams in
Yang-Mills and other supersymmetric theories \cite{Berends:1981rb,Kleiss:1985yh,Xu:1986xb,Gunion:1985vca,Dixon:1996wi,Parke:1986gb,Berends:1987me,
Witten:2003nn,Britto:2004ap,Britto:2005fq,Benincasa:2007xk,Cachazo:2004kj,Georgiou:2004by,Georgiou:2004wu,
Bedford:2005yy,Cachazo:2005ca,ArkaniHamed:2008yf}.
The spinor and twistor
\cite{Witten:2003nn,Cachazo:2004by,Cachazo:2004dr,Mason:2009sa,ArkaniHamed:2009si,Bargheer:2009qu}
representations of the
scattering amplitudes  dramatically simplifies the calculations
\cite{Dixon:1996wi,Britto:2004ap,Britto:2005fq,Benincasa:2007xk}.
The application of the BCFW recurrence relations~\cite{Britto:2004ap,Britto:2005fq} to calculate four-particle amplitudes
allows to derive the production rate of tensor gluons of arbitrary
high spin in the fusion of two gluons $g+g \rightarrow T+T$ \cite{Georgiou:2010mf}.
The consistency of the calculations in different kinematical channels is fulfilled when all
cubic coupling constants between vector bosons (gluons) and high spin tensor bosons
are of the generalized Yang-Mills type \cite{Savvidy:2005fi,Savvidy:2010vb}
and are equal to the Yang-Mills gauge coupling
$
g = g_{YM}.
$
The result can be expressed in a compact form \cite{Georgiou:2010mf}:
\beqa\label{1}
d\sigma_{ g_+g_- \rightarrow T_+T_-}  =
 ~\biggl({1-\cos\theta \over  1+ \cos\theta}\biggr)^{2s-2}
d\sigma_{g_+g_- \rightarrow g_+g_- },  ~~~s=1,2,3,...
\eeqa
where $d\sigma_{g_+g_- \rightarrow g_+g_- }$ is the polarized cross section of two gluons
into two gluons and $\theta$ is the scattering angle.
The formula demonstrates the complete dependence of the cross section on the spin of the  tensor
gluons through the form-factor $F(\theta)^{2s-2}$
\beqa\label{formfactor}
F(\theta)  =
 ~ {1-\cos\theta \over  1+ \cos\theta},
\eeqa
in front of the gluon fusion cross section $d\sigma_{g_+g_- \rightarrow g_+g_- }$.

One can also derive a generalization of the Parke-Taylor scattering amplitude to the case
of two tensor gauge bosons of spin $s$ and $(n-2)$ gluons.
The result of Georgiou-Savvidy reads \cite{Georgiou:2010mf}:
\be\label{GSamplitude}
M_n(1^+,..i^-,...k^{+s},..j^{-s},..n^+)= g^{n-2}
 \frac{<ij>^4}{\prod_{r=1}^{n} <r (r+1)>} \Big( \frac{<ij>}{<ik>}\Big)^{2s-2},
\ee
where $n$ is the total number of particles, and the dots stand for the
positive helicity gluons. Here, $i$ is the position of the negative helicity gluon,
while $k$ and $j$ are the positions of the particles with helicities $+s$ and $-s$ respectively.
This expression is holomorphic in the spinor dependence, exactly as
the MHV gluon amplitude and for $s=1$ the second fraction in \eqref{GSamplitude}
is absent and \eqref{GSamplitude} reduces to the well-known result for
the MHV amplitude \cite{Parke:1986gb}.  It is also remarkable  that the formula
correctly reproduces the tree level gluon scattering amplitudes with the insertion of
quark or scalar pair instead of tensor particles. Indeed this take place when we substitute $s=1/2$ or $s=0$
in (\ref{GSamplitude}).
In a sense the formula has larger validity area than the area in which it has been initially
derived \cite{Georgiou:2010mf}.

The next generalization of  scattering amplitudes for high helicity particles  can be
obtained by considering two pairs of tensor gauge bosons of spin $s_1$ and $s_2$ and $(n-4)$ gluons.
The corresponding amplitude which we have found has the form
\be\label{ASamplitude}
M_n(1^+,..l^{+s_1},.. i^{-s_1},..k^{+s_2},..j^{-s_2},..n^+)= g^{n-2}
 \frac{<ij>^4}{\prod_{r=1}^{n} <r r+1>} \Big( \frac{<ij>}{<ik>}\Big)^{2s_2-2}
 \Big( \frac{<ij>}{<lj>}\Big)^{2s_1-2}
\ee
and is holomorphic in the spinors of the particles,
it reduces to the amplitude (\ref{GSamplitude})
when $s_1 =1$ or when   $s_2 =1$ and to the Parke-Taylor amplitude when both
spins are equal to one.
The derivation of the amplitude (\ref{ASamplitude}) will be presented elsewhere,
here we shall limit ourself in studying  its conformal properties.

It is well known that the MHV Parke-Taylor {\it tree amplitudes} \cite{Parke:1986gb} are
conformal invariant \cite{Witten:2003nn}\footnote{A tree-level holomorphic anomaly
\cite{Cachazo:2004by,Cachazo:2004dr,Mason:2009sa,ArkaniHamed:2009si,Bargheer:2009qu},
caused by the collinear momenta,
can be avoided for tree amplitudes through a choice of external momenta in a general position.}
and that
at the loop-level the conformal symmetry is broken in non-supersymmetric gauge theories like QCD. In $\CN=4$ SYM
the situation is different, it is fully superconformal theory where the redefinition of
the superconformal generators is required in order to restor the  conformal symmetry at the
loop-level \cite{Bargheer:2009qu}.
Our intension is to study the conformal properties of the generalized {\it tree-level} amplitudes (\ref{GSamplitude})
and (\ref{ASamplitude}) which are
describing  the scattering of the tensor gauge bosons. We shall demonstrate that these
generalized {\it tree amplitudes} (\ref{GSamplitude}) and (\ref{ASamplitude}) are conformal invariant\footnote{The interplay
between scale and conformal symmetries in quantum field theory recently were discussed in
a number of publications
\cite{Polchinski:1987dy,Zamolodchikov:1986gt,Cardy:1988cwa,Antoniadis:2011gn,ElShowk:2011gz,
Fortin:2011ks,Fortin:2011sz} }. This is highly unexpected result, as we shell discuses in the
main text of the article that this is not true in particular for the scattering amplitudes of gravitons.
The problem of conformal invariance of higher spin gauge theories is an old and unsolved problem
even at the classical level and the above result can shed some light on that difficult problem.
The higher spin extension of the conformal group have been found recently in \cite{Antoniadis:2011re}, but
it is not yet known if we have here an example of its field theoretical realization.
It is difficult to expect that the scattering amplitudes
of the tensor gauge bosons will remain conformal invariant at the loop-level  and most
probably we shall have the breaking of conformal symmetry similar to the QCD case.

We shall also discuss and compare
the above amplitudes for the  charged tensor gauge bosons with the corresponding amplitudes for
gravitons, stressing their differences and similarities. The differences have their
origin in the charge content of the particles and in the form of the basic cubic interaction vertices
in the corresponding theories, while the similarity lies in the helicity content of the particles.

\section{\it Helicity Amplitudes and Structure of Interaction Vertices}

Let us consider a scattering amplitude for
massless particles of momenta $p_i$ and polarization tensors $\varepsilon_i$ ~$(i=1,...,n)$,
which are described by irreducible massless representations of Poincar\'e group and are
classified by their helicities $h= \pm s$, where $s$ is an integer:
\be\label{smatrix}
M_n = M_n(p_1,\varepsilon_1;~p_2,\varepsilon_2;~...;~p_n,\varepsilon_n).
\ee
We are interested in representing the momenta $p_i$ and polarization tensors $\varepsilon_i$ in terms of spinors
and the above scattering amplitude in terms of rational functions of spinor products
\cite{Berends:1981rb,Kleiss:1985yh,Xu:1986xb,Gunion:1985vca,Dixon:1996wi,Parke:1986gb,Berends:1987me,
Witten:2003nn,Britto:2004ap,Britto:2005fq,Benincasa:2007xk,Cachazo:2004kj,Georgiou:2004by,
Georgiou:2004wu,Bedford:2005yy,Cachazo:2005ca,ArkaniHamed:2008yf}.

The spinor representation of momenta $p^{\mu}$ and
polarization tensors $\varepsilon_i$ can be constructed as follows.
The spinors $\{ \lambda_a, \tilde{\lambda}_{\dot{a}}  \}$ transform in the
representation $(1/2,0)$ and $(0,1/2)$ of the universal cover of the Lorentz
group, $SL(2,C)$, respectively. Invariant tensors are $\epsilon^{ab}$,
$\epsilon^{\dot{a}\dot{b}}$ and $(\sigma^{\mu})_{a\dot{a}}$, where
$\sigma^{\mu}= (1,\vec{\sigma})$. The basic Lorentz invariant spinor products can be
constructed as follows:
$
\lambda_a \lambda'_b \epsilon^{ab}  \equiv <\lambda,\lambda'>,~
\tilde{\lambda}_{\dot{a}} \tilde{\lambda}'_{\dot{b}} \epsilon^{\dot{a}\dot{b}}  \equiv [\lambda,\lambda'].
$
The scalar product of two vectors $p^{\mu}$ and $q^{\nu}$ is given by the product
$
2 (p\cdot q) = <\lambda^p, \lambda^q> [\tilde{\lambda}^p, \tilde{\lambda}^q].
$
Using the third invariant tensor one can define
$p^{\mu} = \lambda^a (\sigma^{\mu})_{a\dot{a}} \tilde{\lambda}^{\dot{a}}$ and find out the
corresponding spinor representation of massless particle momentum in the form
\be\label{momentum}
p_{a\dot{a}}= \lambda_a \tilde{\lambda}_{\dot{a}}.
\ee
The corresponding polarization vectors of spin-1 particles are given by
\be\label{poltensors}
\varepsilon^{-}_{a\dot{a}}(p) = {\lambda_a \tilde{\mu}_{\dot{a}} \over [\tilde{\lambda},\tilde{\mu} ]},~~~
\varepsilon^{+}_{a\dot{a}}(p) = {\mu_a \tilde{\lambda}_{\dot{a}} \over <\mu,\lambda >}
\ee
with $\mu_a$ and $\tilde{\mu}_{\dot{a}}$ as arbitrary reference spinors.
The polarization tensors of massless particles of integer spin $s$ can be expressed in terms
of spin-1 as follows\footnote{ In labeling helicities, we consider all particles to be outgoing.}:
\beqa
\varepsilon^{-}_{a_1\dot{a}_1,...,a_s\dot{a}_s }(p) = \prod^{s}_{i=1} \varepsilon^{-}_{a_i\dot{a}_i},~~~~~~~~
\varepsilon^{+}_{a_1\dot{a}_1,...,a_s\dot{a}_s }(p) = \prod^{s}_{i=1} \varepsilon^{+}_{a_i\dot{a}_i}.
\eeqa
The presence in (\ref{poltensors}) of arbitrary reference spinors
$\mu_a$ and $\tilde{\mu}_{\dot{a}}$ means that polarization tensors
are not uniquely fixed and are changing under the gauge transformations.

The scattering amplitude  of massless bosons $M_n$ (\ref{smatrix}) can  be considered now
as a function of spinors $\lambda_i$, $\tilde{\lambda}_i$ and helicities $h_i$:
$
M_n=M_n(\lambda_1,\tilde{\lambda}_1,h_1;~...;~\lambda_n,\tilde{\lambda}_n,h_n)
$
and  should be a homogeneous function of these spinors of order $2 h_i$
\cite{Witten:2003nn}:
\be\label{helicityequation}
(\lambda^{a}_{i} {\partial \over \lambda^{a}_{i} }  -
\tilde{\lambda}^{a}_{i} {\partial \over \tilde{\lambda}^{a}_{i} })
M_n(...~\{\lambda_i,\tilde{\lambda}_i,h_i\}~...)= -2 h_i ~M_n(...~\{\lambda_i,\tilde{\lambda}_i,h_i\}...).
\ee
From this equation one can derive a general structure of the full three-particle amplitudes $M_3$ in spinor
representation for the {\it complex momenta} \cite{Witten:2003nn,Benincasa:2007xk}.
This  provides a unique information
about the structure of three-point on-shell vertices $M_3(p_1,p_2,p_3)$ in high spin
quantum field theory. Indeed, looking for a polynomial solution  one can find that
the three-point on-shell vertices for complex momenta  have the form\footnote{The earlier investigation of the
three-point vertices in the light-front formulation of relativistic dynamics and
derivation of restrictions imposed on the helicities of scattered particles by the Poincar\'e group
were made in \cite{Bengtsson:1983pd,Bengtsson:1983pg,Bengtsson:1986kh,Metsaev:2005ar}.  In covariant
formulation the interaction vertices were studied in \cite{Berends:1984rq,Berends:1985xx,Berends:1984wp},
see also
\cite{schwinger,fronsdal,Guttenberg:2008qe,fierz,pauli,rarita,singh,singh1,fronsdal1,
Curtright:1987zc,deser}.
The advantage of the spinor formulation is that it gives non-perturbative expressions.}
\cite{Benincasa:2007xk,Georgiou:2010mf}
\beqa\label{threlinearvertex}
M_3(1^{h_1} ,2^{h_2},3^{h_3} ) &=& f <1,2>^{h_3-h_1-h_2} <2,3>^{h_1-h_2-h_3} <3,1>^{h_2-h_3-h_1} + \nn \\
~& &   k ~[1,2]^{-h_3+h_1+h_2} ~[2,3]^{-h_1+h_2+h_3} ~[3,1]^{-h_2+h_3+h_1},
\eeqa
where $f$ and $k$ are momentum independent constants and they are such that if
$
 h_1+ h_2 +h_3 <  0,
$
then  $k=0$, while if
$
 h_1 + h_2 +h_3 >  0
$
then $f=0$ in order to avoid singularities in the limit of the real momenta. Because
the dimensionality of spinors $\lambda$ and
$\tilde{\lambda}$ in formula (\ref{momentum}) is $[mass]^{1/2}$,  the dimensionality
of the three-point on-shell vertex  $M_3$  is $[mass]^{D=\mp(h_1+h_2+h_3)}$.

In the generalized Yang-Mills theory \cite{Savvidy:2005fi,Savvidy:2010vb}
all interaction vertices
between high-spin fields have {\it dimensionless coupling constants in four-dimensional space-time},
therefore one should impose the  constraint $D=\mp(h_1+h_2+h_3)=1$
on the amplitudes $M_3$ in (\ref{threlinearvertex}),
which gives \cite{Georgiou:2010mf}:
\beqa\label{dimensionone1}
M_3 &=& f <1,2>^{-2h_1 -2h_2 -1} <2,3>^{2h_1 +1} <3,1>^{2h_2 +1},~~~~h_3= -1 - h_1 -h_2, \nn\\
~\\
M_3 &=& k [1,2]^{2h_1 +2h_2 -1} [2,3]^{-2h_1 +1} [3,1]^{-2h_2 +1},~~~~~~~~~~~~~~~~~~h_3= 1 - h_1 -h_2.\nn
\eeqa
The formulas  (\ref{dimensionone1})
give a general expression for the dimensionless vertex $M_3$ in terms
of two independent helicities $h_1$ and $h_2$ in the generalized
Yang-Mills theory \cite{Savvidy:2005fi,Savvidy:2010vb}. In particular, one can derive
nontrivial subclass of interaction vertices which contains at least one vector boson of spin 1
and higher-spin  bosons which have
dimensionless coupling constant
\cite{Savvidy:2005fi,Savvidy:2005zm,Savvidy:2005ki,Savvidy:2010vb,Georgiou:2010mf}.
Taking in (\ref{dimensionone1}) $h_1 = \pm 1$ and $h_2 = \pm s$ one can find  $h_3$ from the
equations
$$
h_1+h_2 +h_3 =\pm 1~.
$$
The six corresponding solutions are
$$
h_3 = \pm \vert s-2 \vert,~ \pm s,~
 \pm \vert s+2 \vert~
$$
and the  typical vertex has the form
\beqa\label{1ssvertex}
M^{a_1a_2a_3}_3(1^{-s} ,2^{-1} ,3^{+s} )&=& g ~f^{a_1 a_2 a_3} {<1,2>^{4} \over <1,2> <2,3> <3,1>}
\left({<1,2>  \over  <2,3> }\right)^{2s-2},
\eeqa
and contains a vector and two spin-s bosons.   We have also a vertex which contains spin-$s$ and
spin-$\vert s-2 \vert$ bosons:
\beqa
M^{a_1a_2a_3}_3(1^{-s} ,2^{s-2},3^{+1} )&=& g ~f^{a_1 a_2 a_3} {<1,2>^{4} \over <1,2> <2,3> <3,1>}
\left({<1,3>  \over  <2,3> }\right)^{2s-2},
\eeqa
where $f^{abc}$ are the structure constants of the internal gauge group G.
These vertices reduce to the standard Yang-Mills vertex  when $s=1$ and can be
written in a factorized form
\beqa
M^{a_1a_2a_3 }_3(1^{-s} ,2^{-1} ,3^{+s} )&=& M^{YM~ a_1a_2a_3 }_{3}(1^{-1} ,2^{-1} ,3^{+1} )
\left({<1,2>  \over  <2,3> }\right)^{2s-2},\nn\\
M^{a_1a_2a_3 }_3(1^{-s} ,2^{s-2},3^{+1} )&=& M^{YM~ a_1a_2a_3 }_{3}(1^{-1} ,2^{-1} ,3^{+1} )
\left({<1,3>  \over  <2,3> }\right)^{2s-2}.
\eeqa
The vertex
which does not reduce to the Yang-Mills one contains spin-s and
spin-$\vert s+2 \vert$ bosons and has the form\footnote{Formally it reduces to the
Yang-Mills vertex if one substitutes $s=-1$, sort of analytical continuation of the angular
momentum.  }
\beqa\label{exeptionalvertex}
M^{a_1a_2a_3}_3(1^{-(s+2)} ,2^{s},3^{+1} )&=& g ~f^{a_1 a_2 a_3} {<1,2>^{4} \over <1,2> <2,3> <3,1>}
\left({<1,3>  \over  <2,3> }\right)^{2s+2}.
\eeqa
These are the primitive vertices in the sense that all others can be obtained by the
symmetries such as parity and cyclic $Z_3$ symmetry, which allow  simultaneously
to reverse all helicities in a vertex and  cyclically exchange
their positions. In summary, the above vertices can be written in a
factorized form with a nontrivial prefactor
depending on helicities involved, and
in the general case we can represent the vertex (\ref{dimensionone1})
 in the same factorized form $(h_1+h_2+h_3 =-1)$
\beqa\label{TTV}
M^{a_1a_2a_3}_3(1^{h_1} ,2^{h_2},3^{h_3} ) =
M^{YM~ a_1a_2a_3 }_{3}(1^{-1} ,2^{-1} ,3^{+1} )\left({<2,3> \over  <1,2>  }\right)^{2h_1+2}
 \left({<3,1> \over  <1,2>  }\right)^{2h_2+2}
\eeqa
explicitly exhibiting its dimensionless character.

It is now tempting to present the interaction vertex in gravity where all helicities are
equal to $h_i=\pm 2$ and compare it with the Yang-Mills one
\cite{Kawai:1985xq,Berends:1988zp,Bedford:2005yy,Cachazo:2005ca}.
Using the general formula
(\ref{threlinearvertex}) one can get the cubic vertex for the neutral gravitons:
\beqa\label{gravityvertex}
M^{GR}_3(1^{-2} ,2^{-2} ,3^{+2} )&=& \kappa {<1,2>^{8} \over <1,2>^2 <2,3>^2 <3,1>^2}=
\kappa   \left( M^{YM }_3(1^{-1} ,2^{-1} ,3^{+1} ) \right)^2,~~~
\eeqa
which has dimensionful coupling constant $\kappa^2 = 32\pi G_N$ in front of the quadratic in momenta
spinor expression; one sees that it is exactly equal to the square of the
color/charge  stripped gluon vertex $M^{YM }_3 $. The other vertex
we are interested in is the graviton-photon vertex of helicities $(-2,-1,+1)$
\beqa\label{gravitionphotonvertex}:
M^{MGR}_3(1^{-2} ,2^{-1} ,3^{+1} )&=& \kappa {<1,2>^{4} \over  <2,3>^2  },
\eeqa
which is also quadratic in momentum and has dimensionful coupling constant.

It is
now elucidating  to compare interaction vertices which include neutral gravitons G on the one hand  and charged
non-Abelian tensor bosons T on the other hand.  Both have the same helicities $h=\pm 2$, but the former are neutral while the others are charged. This results in completely different  properties of the interaction
vertices. The graviton vertices include odd number of
gravitons (GGG) (\ref{gravityvertex}) and even number of vector bosons (GVV) (\ref{gravitionphotonvertex})
and the coupling constant is dimensionful, while
the charged tensor bosons vertices include even number of
tensors and odd number of vector bosons (TTV) (\ref{TTV}) with dimensionless coupling constant.
The differences will become more transparent  when we compare the tree level amplitudes in the corresponding
theories in the next sections.

\section{\it High Spin Helicity Amplitudes}

With these vertices in hand one can compute the gluon fusion amplitudes into two
high spin-s gauge bosons $s=2,3,...$.
For the four-particle scattering amplitudes one can find the
following expression \cite{Georgiou:2010mf}:
\beqa\label{firstseconddiagramgeneralholomorphicf}
M^{abcd}_4(1^+ ,2^-, 3^{+s},4^{-s})
 = -2 i  g^2  \left({ <2,4>  \over <2,3>    }\right)^{2s-2}
  \times  ~~~~~~~~~~~~~~~~~~~~~~~~\\
  \nn\\
~~~~~~~~~~~~~~~~~~~~~ \left( f^{ade} f^{bce}   { <2,4>^4  \over  \prod <i,i+1>  }
 +f^{ace} f^{bde}   {<2,4>^4  \over  \prod_{3\leftrightarrow 4} <i,i+1> } \right) ,\nn
\eeqa
where in the brackets of the second line above is the gluonic amplitude $g_++g_- \rightarrow g_++g_-$
times the form-factor
which is the contribution of the high spin gauge bosons.
This is  {\it purely holomorphic expression}.
The alternative helicity amplitude  $M^{abcd}_4(1^+,1^-, 3^{-s},4^{+s})$ can
also be found \cite{Georgiou:2010mf}.
Here the four gluon scattering amplitudes without color factors in (\ref{firstseconddiagramgeneralholomorphicf})
(the color ordered amplitudes) have the form
\beqa\label{yangmillsamplitude}
M^{YM}_4(1^{+1},2^{-1},3^{+1},4^{-1}) &=&  {<2,4>^{4} \over \prod <i,i+1>  },~~\nn\\
M^{YM}_4(1^{+1},2^{-1},4^{-1},3^{+1}) &=&  {<2,4>^{4} \over \prod_{3\leftrightarrow 4} <i,i+1>  }.
\eeqa

We shall also present the four graviton scattering amplitude
\cite{Kawai:1985xq,Berends:1988zp,Bedford:2005yy,Cachazo:2005ca}
\beqa\label{fourpointamplitudegravity}
M^{GR}_4(1^{+2},2^{-2},3^{+2},4^{-2})=
   \frac{ [34] <24>^7 }{<12>^2 <23>  <34>  <41> <3,1>} ,
\eeqa
which  can be expressed in terms of four gluon scattering amplitudes (\ref{yangmillsamplitude})
\beqa\label{fourpointamplitudegravitygluons}
M^{GR}_4(1^{+2},2^{-2},3^{+2},4^{-2})= -s_{12} M^{YM}_4(1^{+1},2^{-1},3^{+1},4^{-1})   \times
  M^{YM}_4(1^{+1},2^{-1},4^{-1},3^{+1}),~~~~~~~
\eeqa
but with an additional energy factor $s_{12} = (p_1+p_2)^2 = (p_3+p_4)^2 $. These relations
can be easily understood on dimensional ground: in gauge theory the vertex is linear in
 momenta while in gravity it is quadratic, therefore one should have an energy factor in front
 of the gauge theory amplitudes.

The next interesting result can be obtained if one considers the color-ordered
scattering amplitudes involving two tensor particles of  helicity $+s$ and $-s$
respectively, one negative helicity gluon
and any number of gluons with positive helicity. In the case when $s=1$ one has
the MHV amplitude  for the scattering of vector bosons (gluons).
The expression for this amplitude is given by the
famous Parke-Taylor formula \cite{Parke:1986gb}
\be\label{park}
\hat{M}_n(1^+,...i^-, ... , j^{- },...,n^+)=
i g^{n-2} (2\pi)^4 \delta^{(4)}(P^{a\dot{b}}) \frac{<ij>^4}{\prod_{l=1}^{n} <l l+1>}  ,
\ee
where
$$
P^{a\dot{b}} = \sum^n_{m=1} \lambda^a_m \tilde{\lambda}^{\dot{b}}_m
$$  is the total momentum.

The generalization of the above formula to the case of two spin $s$ bosons
and $(n-2)$ gluons has been found  in  \cite{Georgiou:2010mf}:
\be\label{gs}
\hat{M}_n(1^+,..i^-,...k^{+s},..j^{-s},..n^+)=i g^{n-2} (2\pi)^4 \delta^{(4)}(P^{a\dot{b}})
 \frac{<ij>^4}{\prod_{l=1}^{n} <l l+1>} \Big( \frac{<ij>}{<ik>}\Big)^{2s-2},
\ee
where $n$ is the total number of particles, and the dots stand for any number of
positive helicity gluons. Finally, $i$ is the position of the negative helicity gluon,
while $k$ and $j$ are the positions of the particles with helicities $+s$ and $-s$ respectively.
The first comment is that the expression \eqref{gs} is holomorphic in the spinors of the particles,
exactly as the MHV gluon amplitude \eqref{park} is.
The second comment is that for $s=1$ the second fraction in \eqref{gs}
is absent and \eqref{gs} reduces to the  MHV amplitude \eqref{park}.
In particular the five-particle amplitude takes the following form\footnote{The 4- and 5-particle amplitudes
correspond to 2-jet and 3-jet production in hadronic
collisions at very high energies, and so are of phenomenological interest.}:
\beqa\label{fivepointamplitude}
\hat{M}_5(1^+,2^-,3^{+},4^{+s},5^{-s})=  i g^{3} (2\pi)^4 \delta^{(4)}(P^{a\dot{b}})
\frac{<25>^4}{\prod_{i=1}^{5}
<i i+1>}  (\frac{<25> }{<24>})^{2s-2} .
\eeqa
The third comment is that the ratio of the amplitudes for the pair of tensor gauge bosons of spin $s_1$ and
spin $s_2$ has the form
\be
\hat{M}_n(1^+,..i^-,...k^{+s_1},..j^{-s_1},..n^+) = \Big( \frac{<ij>}{<ik>}\Big)^{2s_1-2s_2}
\hat{M}_n(1^+,..i^-,...k^{+s_2},..j^{-s_2},..n^+)
\ee
and is in formal agrement with the supersymmetric Ward identities
\cite{Grisaru:1976vm,Grisaru:1977px,Dixon:1996wi}
connecting gluon or graviton amplitudes
with the insertion of a pair of partons $P$
\be
A^{SUSY}_n (1^+,..i^-,...k^{+h_P},..j^{-h_P},..n^+)=
\Big( \frac{<ij>}{<ik>}\Big)^{2h_P-2h_{\phi}} A^{SUSY}_n (1^+,..i^-,...k^{+ }_{\phi},..j^{-  }_{\phi},..n^+),
\ee
where $P$ refers to a scalar, fermion, gluon, gravitino or graviton with respective helicity
$h_P=0,1/2,1,3/2,2$ and $\phi$ refers to a scalar particle, or graviton.

The other generalization of the scattering amplitudes for high helicity particles  can be
obtained by considering a pair of particles of spin $s_1$ and $s_2$ and (n-4) gluons.
The corresponding amplitude has the form
\beqa\label{AS}
M_n(1^+,..l^{+s_1},.. i^{-s_1},..k^{+s_2},..j^{-s_2},..n^+)=
~~~~~~~~~~~~~~~~~~~~~~~~~~~~~~~~~~~~~~~~~~\nn\\
\\
~~~~~~~~~~~~~~~~ =i g^{n-2} (2\pi)^4 \delta^{(4)}(P^{a\dot{b}}) \frac{<ij>^4}{\prod_{r=1}^{n} <r r+1>} \Big( \frac{<ij>}{<ik>}\Big)^{2s_2-2}
 \Big( \frac{<ij>}{<lj>}\Big)^{2s_1-2}\nn
\eeqa
and is holomorphic in the spinor dependence, reduces to the amplitude (\ref{gs})
when $s_1 =1$ or $s_2 =1$ and reduces to the Parke-Taylor amplitude when both spins are equal to one.

The exceptionally interesting non-MHV amplitude which involves two vector bosons and two tensor bosons
of  different spins  $s$ and $s-2$ is \cite{Georgiou:2010mf}:
\beqa\label{seconddiagramsdifs}
 M^{abcd}_4(+1,-s, +1,s-2) =
 -2 i  g^2  \left({ <2,3>  \over <3,4>    }\right)^{2s-2}
  \times  ~~~~~~~~~~~~~~~~~~~~~~~~\\
  \nn\\
~~~~~~~~~~~~~~~~~~~~~ \left( f^{ade} f^{bce}   { <2,4>^4  \over  \prod <i,i+1>  }
 +f^{ace} f^{bde}   {<2,4>^4  \over  \prod_{3\leftrightarrow 4} <i,i+1> } \right) ,\nn
\eeqa
where  $s=2,3,...$. It is interesting to notice that if one formally substitutes $s=1$ into the above expression, then
one can see that it reduces to the correct Yang-Mills amplitude $ M^{abcd}_4(1,-1, 1,-1)$, as
it already happened in the case of the corresponding three-point on-shell vertex (\ref{exeptionalvertex}).

This amplitude is holomorphic and is
of special interest because it has only {\it one particle of negative helicity}.
In comparison, the $n$-gluon
tree amplitudes for $n \geq 4$ with all but one gluon of
positive helicity vanish \cite{Parke:1986gb,Witten:2003nn}.
Thus the $n$-particle tree amplitudes in the generalized Yang-Mills theory have more complicated structure.
It is well known that the tree level $n$-gluon scattering
amplitudes with all positive helicities also vanish \cite{Parke:1986gb,Witten:2003nn}.
In generalized Yang-Mills theory this statement remains true and can be
proved by induction \cite{Georgiou:2010mf}.   {\it Thus in the generalized Yang-Mills
theory the tree level n-particle
scattering amplitudes with all positive helicities vanish $M^{tree}_n(+,...,+)=0$,
but tree amplitudes with one negative
helicity particle are already nonzero} (\ref{seconddiagramsdifs}).

Our intension in the next section is to check the conformal properties of the
amplitudes (\ref{gs}) and (\ref{AS}).

\section{\it Conformal Invariance of Tensor Bosons MHV Amplitudes}

We are interested now in studying  the conformal properties of the above generalized MHV amplitudes
\eqref{gs}. The generators of the conformal algebra expressed in terms of the $\lambda$
and $\tilde{\lambda}$ variables are given in \cite{Witten:2003nn} and have the following form:
\beqa
P_{a\dot{b}}= \lambda_a \tilde{\lambda}_{\dot{a}} ,~~~~  J_{ab}
= {i \over 2} (\lambda_a {\partial \over \partial \lambda^b} +
\lambda_b {\partial \over \partial \lambda^a})  ,~~~~
\tilde{J}_{\dot{a}\dot{b}} = {i \over 2} (\tilde{\lambda}_{\dot{a}} {\partial \over \partial \tilde{\lambda}^{\dot{b}}}
+ \tilde{\lambda}_{\dot{b}} {\partial \over \partial \tilde{\lambda}^{\dot{a}}})\nn\\
D= {i \over 2} (\lambda^a {\partial \over \partial \lambda^a} +
\tilde{\lambda}^{\dot{a}} {\partial \over \partial \tilde{\lambda}^{\dot{a}}} +2),~~~~~~
K_{a\dot{b}}= {\partial^2 \over \partial \lambda^a  \partial  \tilde{\lambda}^{\dot{b}}}~~~~~~~~~~~~~~~~~
\eeqa
Lorentz invariance of the formula  \eqref{gs}  is manifest, the momentum conservation is
fulfilled
because of the delta function and we have to verify that the amplitude is
annihilated by D and by K generators.

The dilatation operator D annihilates MHV amplitude \eqref{park}  \cite{Witten:2003nn},
therefore one should check that the additional terms arising from the factor
\be\label{formfactor}
\Big( \frac{<ij>}{<ik>}\Big)^{2s-2}
\ee
in the  amplitude \eqref{gs}  also vanish. The derivatives
acting on the spinors $\lambda_j$ and $\lambda_k$ cancel each other and the derivative
over the spinor $\lambda_i$ vanishes as well, therefore $D\hat{ M}_n =0$.

Similarly the special conformal generator
$K_{a\dot{a}}$ annihilates the MHV  amplitude \eqref{park}  \cite{Witten:2003nn},
and one should check that it annihilates the amplitude in the presence of the
form-factor \eqref{formfactor}.
Since $\partial M_n / \partial  \tilde{\lambda} =0$ and $J_{ab} M_n =0$ the main
part of the calculations in \cite{Witten:2003nn} remains valid and we have
\beqa\label{conformalgenerator}
K_{a\dot{a}} \hat{M}_n &=&{\partial^2 \over \partial \lambda^a  \partial  \tilde{\lambda}^{\dot{a}}}
\hat{M}_n    \\
&=& i g^{n-2} (2\pi)^4 \left( (n-4) M_n ~{\partial \over \partial P^{a\dot{a}} }~
\delta^{(4)}(P) + ~(\sum_m   \lambda^b_m ~{\partial M_n \over \partial \lambda^a_m}~)~
 {\partial \over \partial P^{b \dot{a}} }~ \delta^{(4)}(P)  \right).\nn
\eeqa
The last sum can be evaluated further:
$$
\sum_m  ~ \lambda^b_m ~{\partial M_n \over \partial \lambda^a_m} =
{1 \over 2}~ \delta^{b}_{a} ~\sum_m   \lambda^c_m {\partial M_n \over \partial \lambda^c_m},
$$
where
$$
\sum^n_{m=1}   \lambda^c_m {\partial M_n \over \partial \lambda^c_m} = (-2(n-3) +2 -2s +2s)M_n=
-2(n-4)M_n,
$$
and plugging these expressions back into \eqref{conformalgenerator} yields
$K_{a\dot{a}} \hat{M}_n =0$. Similar calculation for the amplitude (\ref{AS}) gives
$$
\sum^n_{m=1}   \lambda^c_m {\partial M_n \over \partial \lambda^c_m} = (-2(n-4) +2s_1 -2s_1 +2s_2 -2s_2)M_n=
-2(n-4)M_n.
$$
{\it Thus the generalized tree amplitudes  \eqref{gs} and \eqref{AS} are conformally
invariant.} As it is easy to check the four graviton scattering amplitude (\ref{fourpointamplitudegravity})
is not conformally invariant because of the spinor product $[1,3]$ in (\ref{fourpointamplitudegravity}).
This is another manifestation of the difference between charge and neutral helicity-two particles
which has its origin in the structure of the corresponding vertices   (\ref{TTV}),
(\ref{gravityvertex}) and  (\ref{gravitionphotonvertex}).

We do not  know whether {\it all tree level amplitudes} of high spin gauge theory are conformally invariant
or not. The reason is that presently there are very limited higher spin amplitudes
which are available for the analyzes. The on-shell scattering amplitudes  \eqref{gs} and \eqref{AS}
became available only thanks to the new recursion relations for the tree-level amplitudes
for the lower spin gauge theories set up in
\cite{Berends:1981rb,Kleiss:1985yh,Xu:1986xb,Gunion:1985vca,Dixon:1996wi,Parke:1986gb,Berends:1987me,
Witten:2003nn,Britto:2004ap,Britto:2005fq,Benincasa:2007xk,Cachazo:2004kj,Georgiou:2004by,Georgiou:2004wu,
Bedford:2005yy,Cachazo:2005ca,ArkaniHamed:2008yf} .
The problem of conformal invariance of higher spin gauge theories is an old and unsolved problem
even at the classical level and the above result can shed some light on that difficult problem.
The higher spin extension of the conformal group have been found recently in \cite{Antoniadis:2011re}, but
it is not yet known if we have here its field theoretical realization.
The progress in the construction of multi-particle and multi-loop scattering amplitudes
in $\CN = 4$ SYM in spinor and twistor spaces
\cite{Nair:1988bq,Drummond:2008cr,Drummond:2009fd,Mason:2009sa,ArkaniHamed:2009si,Drummond:2008vq}
raises a hope that future progress can be reached also in higher spin gauge theories.
It is difficult to expect that the scattering amplitudes
of the tensor gauge bosons will remain conformal invariant at the loop-level  and most
probably we shall have the breaking of conformal symmetry similar to the QCD case.

We have to notice also that whenever two momenta become collinear, conformal
symmetry becomes anomalous \cite{Cachazo:2004by,Cachazo:2004dr,Mason:2009sa,ArkaniHamed:2009si}
and the naive action of infinitesimal conformal transformations on
scattering amplitudes is not complete \cite{Mason:2009sa,ArkaniHamed:2009si,Bargheer:2009qu}.  It needs to be supplemented by correction terms
which cure the collinear anomaly at tree level. The corrections have the ability to change the number
of legs of the scattering amplitudes, so that such generators act on the generating functional
of all amplitudes \cite{Bargheer:2009qu}.
In the next section we shall study collinear momenta behavior of the scattering amplitudes
of tensor gauge bosons and shall compare it with the YM case.

\section{\it Collinear Behavior and Splitting Functions}

The color-ordered amplitudes in QCD can only have poles in channels of
cyclically adjacent momenta \cite{Dixon:1996wi}. But, in fact, only in two-particle
channels the corresponding MHV amplitudes develop the poles, due to the vanishing
of  $M^{tree}_n(1^{\pm}, 2^+,...,n^+)$ amplitudes. Thus only collinear (two-particle) singularities
of adjacent particles are permitted  and the collinear behavior
of the tree amplitudes has the following general form
\cite{Dixon:1996wi,Parke:1986gb,Berends:1987me,Berends:1988zn,Mangano:1987kp}:
\be
M^{tree}_n(...,a^{\lambda_a},b^{\lambda_b},...)~~  {a \parallel b \over \rightarrow}  ~~  \sum_{\lambda=\pm 1}
Split^{ tree }_{-\lambda}(a^{\lambda_a},b^{\lambda_b})~ \times ~M^{tree}_{n-1}(...,P^{\lambda} ,...),
\ee
where $Split^{tree}_{-\lambda}(a^{\lambda_a},b^{\lambda_b})$ denotes the splitting amplitude, the
intermediate state $P$ has momentum $k_P=k_a +k_b$ and helicity $\lambda$.

In the generalized Yang-Mills theory the tree amplitudes with only one negative helicity
are not vanishing $M^{tree}_{n}(-,+,...,+) \neq 0$, as we have seen at the end
of the third section (\ref{seconddiagramsdifs}). Therefore multi-particle pole
can appear in the corresponding limit, if tensor bosons have different spins: let us say,
$s$ and $s-2$. Bellow we shall consider only the case of equal spins. An easy way to extract the splitting
amplitudes in that case will be to consider five-point amplitude \eqref{fivepointamplitude}.
Let us consider the amplitude \eqref{fivepointamplitude} in the limit when the
particles 4 and 5 become collinear: $k_4  \parallel k_5$, that is,
$k_4 = x k_P,~k_5 = (1-x) k_P$,  $k^2_P \rightarrow 0$ and $x$ describes the
longitudinal momentum sharing. In this limit
we have
$$
\lambda_4 = \sqrt{x} \lambda_P,~~~\lambda_5 = \sqrt{1-x} \lambda_P,
$$
and can bring the amplitude in the following form:
\beqa
M_5(1^+,2^-,3^{+},4^{+s},5^{-s})&=& \frac{<2P>^4}{ <1, 2><2, 3><3, P> <P, 1>}
  \frac{(1-x)^{ s+1/2} }{x^{s-1/2} } \frac{1}{ <4, 5>}=\nn\\
  \\
  &=& A_4(1^+,2^-,3^{+},P^-) \times ~ Split_+(a^{+s},b^{-s}),\nn
\eeqa
where
\be
Split_+(a^{+s},b^{-s}) = \left(\frac{1-x}{x } \right)^{ s-1}  \frac{(1-x)^2}{\sqrt{x(1-x)}}
\frac{1}{ <a, b>}
\ee
and one can deduce that
\be
Split_+(a^{-s},b^{+s}) = \left(\frac{x }{1-x} \right)^{ s-1}  \frac{x^2}{\sqrt{x(1-x)}}
\frac{1}{ <a, b>}.
\ee

Considering different collinear limits $k_1  \parallel k_5$ and $k_3  \parallel k_4$
one can get
\beqa
Split_{+s}(a^{+},b^{-s}) &=&  (1-x)^{ s-1}  \frac{(1-x)^2}{\sqrt{x(1-x)}}
\frac{1}{ <a, b>},\nn\\
Split_{+s}(a^{-s},b^{+}) &=&  x^{ s-1}  \frac{x^2}{\sqrt{x(1-x)}}
\frac{1}{ <a, b>}
\eeqa
and
\beqa
Split_{-s}(a^{+s},b^{+}) &=&  \left(\frac{1} {x} \right)^{ s-1}  \frac{1}{\sqrt{x(1-x)}}
\frac{1}{ <a, b>},\nn\\
Split_{-s}(a^{+},b^{+s}) &=&  \left(\frac{1} {1-x } \right)^{ s-1}  \frac{1}{\sqrt{x(1-x)}}
\frac{1}{ <a, b>}.
\eeqa
This set of splitting functions for the amplitudes $V \rightarrow TT$ and $T \rightarrow VT$
reduces to the gluon splitting functions
\cite{Dixon:1996wi,Parke:1986gb,Berends:1987me,Berends:1988zn,Mangano:1987kp} if one takes $s=1$.
The residue of the collinear pole in the square of the factorized amplitude gives the probability
of creating a pair of tensor gauge bosons
\be
\sigma^{ng TT}  = \sigma^{(n-1)g}  ~\times~ P_{gTT}(x),
\ee
where
\be
P_{gTT}(x) =  \left( {x^2 \over \sqrt{x(1-x)}}\right)^{2}  \left( {x\over 1-x} \right)^{2s-2} +
 \left( {(1-x)^2 \over \sqrt{x(1-x)}} \right)^{2}   \left( {1-x\over  x}\right)^{2s-2},
\ee
with strong amplification of the forward and backward creation probability.
We shall consider the multi-particle pole structure in a future work.

\section*{\it Acknowledgement}
We would like to thank  L.~Brink, L.~J.~Dixon,  J.~M.~Drummond  and T.~R.~Taylor for helpful discussions.
One of us, G.S., would like to thank the CERN Theory Division for hospitality and G. Georgiou for
discussions and participation in the early stages of this work.
This work was supported in part by the European Commission under the ERC Advanced
Grant 226371 and the contract PITN-GA-2009-237920.

\end{document}